\documentclass[a4paper]{article}
\input epsf

\begin {document} 

\title {\bf TRAJECTORIES IN THE CONTEXT OF THE QUANTUM NEWTON'S LAW }
\author{A.~Bouda\footnote{Electronic address: 
{\tt bouda\_a@yahoo.fr}} \ and T.~Djama\footnote{Electronic address: 
{\tt djama.toufik@caramail.com}}\\
Laboratoire de Physique Th\'eorique, Universit\'e de B\'eja\"\i a,\\ 
Route Targa Ouazemour, 06000 B\'eja\"\i a, Algeria\\}

\date{\today}

\maketitle

\begin{abstract}
\noindent
In this paper, we apply the one dimensional quantum law of motion, 
that we recently formulated in the context of the trajectory 
representation of quantum mechanics, to the constant potential, the linear 
potential and the harmonic oscillator. In the classically 
allowed regions, we show that to each classical trajectory there 
is a family of quantum trajectories which all pass through some 
points constituting nodes and belonging to the classical 
trajectory. We also discuss the generalization to any potential 
and give a new definition for de Broglie's wavelength in such a way 
as to link it with the length separating  adjacent nodes. 
In particular, we show how quantum trajectories have as a limit 
when $\hbar \to 0$ the classical ones. In the classically forbidden 
regions, the nodal structure of the trajectories is lost and the 
particle velocity rapidly diverges. 
 
\end{abstract}

\vskip\baselineskip



\newpage

\vskip0.5\baselineskip
\noindent
{\bf 1.\ \ Introduction }
\vskip0.5\baselineskip

For a one-dimensional system of energy $E$ and potential $V(x)$, 
the quantum stationary Hamilton-Jacobi equation (QSHJE) is
\begin {equation}
{1\over 2m} \left({\partial S_0 \over \partial x}\right)^2 + V(x)-E= 
{\hbar^2\over 4m}  \left[{3\over 2}\left( 
{\partial S_0 \over\partial x}\right)
^{- 2 }\left({\partial^2 S_0 \over \partial x^2}\right)^2-
\left( {\partial S_0 \over \partial x  }\right)^{- 1 }
\left({\partial^3 S_0 \over \partial x^3  }\right) \right] \; .
\end {equation}
The solution of this equation is investigated in Refs. 
\cite{FM1, FM2, FM3, B1, F34, F9, F214, F00}. It is shown 
that it can be written as \cite{BD}  
\begin {equation}
S_0=\hbar \ \arctan {\left [ a  {\phi_1 
\over \phi_2 } +b \right ]} +\hbar l \; ,
\end {equation}
where $(\phi_1,\phi_2)$ is a set of two real independent solutions
of the Schr\"odinger equation  
\begin {equation}
-{\hbar^2 \over 2m} {d^2 \phi \over dx^2 } + V(x) \phi = E \phi 
\end {equation}
and \ $(a,b,l)$ \ are real integration constants satisfying 
the condition $a \not= 0$. In Eq. (2), $S_0$ depends also on 
the energy $E$ through the solutions $\phi_1$ and $\phi_2$.
Remark that both for the classical stationary Hamilton-Jacobi 
equation and the QSHJE, if $S_0$ is solution, $-S_0$ is also 
solution. Therefore, the conjugate momentum is given by 
\begin{equation}
P =  {\partial S_0\over\partial x}
  =  \pm {\hbar a W \over {\phi_2^2 + 
         (a\phi_1 + b\phi_2)^2}} \; , 
\end {equation}
where $W = \phi_{1}' \phi_2 - \phi_1 \phi_{2}'$ is a constant 
representing the Wronskian of $(\phi_1, \phi_2)$. As also observed in 
\cite{F9,F14,F13,BFM}, the $\pm$ sign in Eq. (4) indicates that 
the motion may be in either direction on the $x$ axis.
In contrast with Bohm's theory, it is shown in Refs. 
\cite{FM1,FM2,FM3,B1,BFM} that it is possible to relate the 
reduced action $S_0$ to the Schr\"odinger wave function in a 
unified form both for bound and unbound states so that
the conjugate momentum never has a vanishing value. 

Recently \cite{BD}, by taking advantage of the fact that the 
solution of (1) is known, we constructed a Lagrangian from 
which we derived the fundamental relation
\begin {equation}
\dot{x}{\partial S_0 \over \partial x}=
2[E-V(x)] \; .
\end {equation}
By using (4) in this last equation, we get
\begin {equation}
{dx \over dt}= \pm { 2[E-V(x)] \over \hbar a W}
\left[\phi_2^2 + (a \phi_1 + b \phi_2)^2 \right] \; .
\end {equation}
In what follows, we adopt the following convention: the sign 
of the parameter $a$ is chosen so that $aW > 0$. In this way, 
if the particle moves in the classically allowed region 
$(E>V)$ in the positive direction, we must 
use the plus sign in the right hand side (RHS) of (6). When the 
particle gets to a turning point, we must use the minus sign 
whether the particle remains in the classically allowed region 
by changing its direction of motion or it enters the 
classically forbidden region $(E<V)$ by keeping its direction of 
motion. In other words, when the particle changes the branch on 
its trajectory at the turning point, even if it passes to the 
classically forbidden region, the sign which precedes the RHS 
of (6) must be changed.

In Ref. \cite {BD}, we showed that relation (5) leads to 
a third order differential equation representing the first 
integral of the quantum Newton's law (FIQNL)
\begin {eqnarray}
(E-V)^4-{m{\dot{x}}^2 \over 2}(E-V)^3+{{\hbar}^2 \over 8} 
{\left[{3 \over 2}
{\left({\ddot{x} \over \dot{x}}\right)}^2-{\dot{\ddot{x}} \over \dot{x}} 
\right]} (E-V)^2\hskip15mm&& \nonumber\\
-{{\hbar}^2\over 8}{\left[{\dot{x}}^2 
{d^2 V\over dx^2}+{\ddot{x}}{dV \over dx}
 \right]}(E-V)-{3{\hbar}^2\over 16}{\left[\dot{x}
{dV \over dx}\right]^2}=0 \; .
\end {eqnarray}
The solution $x(t)$ of this equation will contain the two 
usual integration constants $E$ and $x_0$ and two additional 
constants which we will call the non-classical integration 
constants. All these constants can be determined by 
the knowledge of $x(t_0)$, $\dot{x}(t_0)$, $\ddot{x}(t_0)$ 
and $\dot{\ddot{x}}(t_0)$.

Without appealing to the Lagrangian formulation, we emphasize that 
relation (5)

- is obtained by using the quantum version of Jacobi's theorem \cite{BD};

- can be obtained by the Hamiltonian formulation.

In this paper, we apply respectively in Sections 2, 3 and 4 the 
quantum law of motion (5) or (7) in the cases of a constant 
potential, a linear potential and a harmonic oscillator. In 
Section 5, we comment on the generalization to any potential 
of the obtained results and give a new definition 
for de Broglie's wavelength and its physical meaning in 
trajectory interpretation of quantum mechanics.

\vskip0.5\baselineskip
\noindent
{\bf 2.\ \ Constant potential }
\vskip0.5\baselineskip

Let us consider the case in which the potential is constant 
$V(x)=V_0$ and set
\begin {equation}
\epsilon = E-V_0 \; .
\end {equation}

We begin by the classically allowed case $(\epsilon > 0)$. 
With the same procedure which we have used in Ref. \cite{BD} for 
the free particle, we can integrate (7) after having 
substituted $V(x)$ by $V_0$. We obtain
\begin {equation}
x(t)=\pm {\hbar\over\sqrt{2m \epsilon}} \arctan{\left[a 
\tan{\left({2 \epsilon t \over\hbar}\right)}+b\right]}+x_0  \; .
\end {equation}
Note that for the particular values $a=1$ and $b=0$ of the 
non-classical integration constants, expression (9) reduces to the
classical relation
$$
x(t)= \pm \sqrt{2 \epsilon \over m}\;t+x_0 \; 
$$
whether the velocity is positive or negative. 

Since the arctangent function is contained between $-\pi /2$ and 
$\pi /2$, it is necessary to readjust the additive integration 
constant $x_0$ after every interval of time in which the tangent 
function goes from $-\infty$ to $+\infty$. This readjustment 
must be made in such a way as to guarantee the continuity 
of $x(t)$. For this reason, expression (9) 
must be rewritten as follows 
\begin {equation}
x(t)={\hbar\over\sqrt{2m \epsilon}} \arctan{\left[a 
\tan{\left({2 \epsilon t \over\hbar}\right)}+b\right]}+
{\pi \hbar \over \sqrt{2m \epsilon} }n+x_0  
\end {equation}
with 
$$
t \in \left \lbrack {\pi \hbar \over 2\epsilon }\left(n-{1 \over 2 }\right),
 {\pi \hbar \over 2\epsilon }\left(n+{1 \over 2 }\right)   
 \right \rbrack
$$
for every integer number $n$. In (10), the parameter $a$ is 
assumed positive and we have considered only the case of 
positive velocity.
In Fig. 1, we have plotted in  $(t,x)$ plane for a free 
electron of energy $E=10$ eV some trajectories corresponding 
to different values of $a$ and $b$. All these trajectories, 
even the classical one $(a=1, b=0)$, pass through some points 
which we will call nodes and which correspond to the times 
\begin {equation}
t_n = {\pi \hbar \over 2\epsilon } \left(n+{1 \over 2}\right)
\end {equation}
for which $ x(t)$ does not depend on $a$ and $b$. The 
distances between two adjacent nodes on the time axis
\begin {equation}
\Delta t_n = t_{n+1}-t_n = {\pi \hbar \over 2 \epsilon}
\end {equation}
and the space axis
\begin {equation}
\Delta x_n = x(t_{n+1})-x(t_n) = 
{\pi \hbar \over \sqrt{ 2m \epsilon}}
\end {equation}
are both proportional to $\hbar$. This means that in the classical 
limit $\hbar \to 0$, the nodes become infinitely close and, then, all 
possible quantum trajectories tend to be identical to the classical 
one. In fact, let us consider an arbitrary point $M(t_M,x_M)$ on 
any quantum trajectory between two adjacent nodes 
$(t_{n-1},x_{n-1})$ and $(t_n,x_n)$. Considering that the variable 
on the $t$ axis is a product of a unit velocity by the time, the 
distance between $M$ and its orthogonal projection  
$M_0(t_{M_0},x_{M_0})$ on the classical trajectory is 
\begin {equation}
MM_0 = \sqrt{{m \over 2\epsilon} +1} \; |t_M - t_{M_0}| \; .
\end {equation}
Note that this relation can be obtained without using the 
expression for $x(t)$ corresponding to the trajectory on which 
$M$ is located. From Eq. (6), we can see that for any 
potential and in any interval which does not contain turning 
points, the function $x(t)$ is monotonous. In the case of 
Fig. 1, the function $x(t)$ is increasing and, then, we have 
$t_{n-1} \leq t_M  \leq t_n$
and
$t_{n-1} \leq t_{M_0} \leq t_n$.
This implies that $|t_M - t_{M_0}| \leq t_n - t_{n-1}$ and, with 
the use of (14), it follows that $MM_0 \to 0$ in the limit 
$\hbar \to 0$. Of course, if $x(t)$ is a decreasing function, 
we also get to the same conclusion. 
This is the fundamental reason why in problems for which the 
constant $\hbar$ can be disregarded, quantum trajectories 
reduces to the classical one. This conclusion is not compatible  
with the finding of Floyd \cite{F15} who states that a residual 
indeterminacy subsists when we take the classical limit. It is not 
also compatible with our previous paper \cite{BD} in which we have 
not taken into account the presence of these nodes.

Finally, note that the solution (9) of (7) in the case where 
$V(x) = V_0$  can be also obtained from the differential 
equation (6),
\begin {equation}
{dx \over dt} = \pm {1 \over a}\sqrt{{2\epsilon \over m}} 
\left[ \cos^2 \left({\sqrt{2m\epsilon} \over \hbar }x \right) +
 \left[a \, \sin \left({\sqrt{2m\epsilon}\over \hbar }x \right) + 
b \, \cos \left({\sqrt{2m\epsilon}\over \hbar }x \right)\right]^2 
    \right], 
\end {equation}
in which we have chosen as solutions of the Schr\"odinger 
equation, Eq. (3), the functions 
$\phi_1=\sin(\sqrt{2m\epsilon}\; x / \hbar)$
and
$\phi_2=\cos(\sqrt{2m\epsilon}\; x / \hbar)$.

Now, let us consider the classically forbidden case $(\epsilon < 0)$. 
Eq. (7) takes the form

\begin{figure}
\def\put(#1,#2)#3{\leavevmode\rlap{\hskip#1\unitlength\raise#2\unitlength\hbox{#3}}}
\centerline{
\vbox{\hsize=10.5cm
\setlength{\unitlength}{1truecm}
\put(0,0){\epsfxsize=10cm \epsfbox{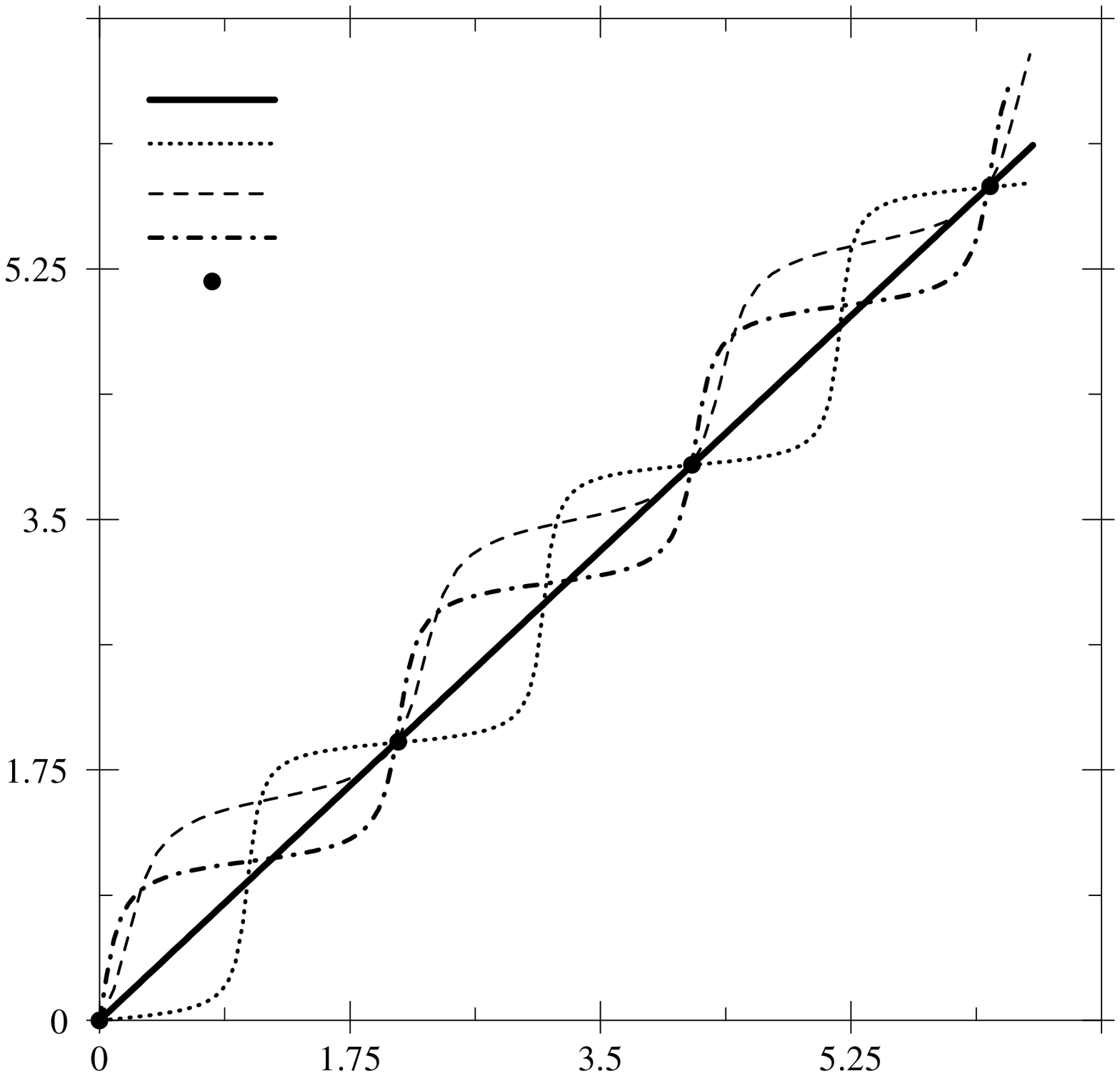}}%
\put(8.5,0.6) {$t$ ($\times 10^{-16}$ s)}%
\put(0.2,9.75) {$x$ ($\times 10^{-10}$ m)}%
\put(2.8,8.8) {\footnotesize  Classical trajectory $(a=1,b=0)$}%
\put(2.8,8.4) {\footnotesize $ a=10 ,\ b=0$}%
\put(2.68,8.) { \footnotesize $ a=3 ,\ b=2$}%
\put(2.68,7.6) { \footnotesize $ a=1/2 ,\  b=1.5$}%
\put(2.68,7.2) { \footnotesize Nodes}%
}}
\centerline {
\vbox 
{ \hsize=12cm\noindent 
\small Fig. 1:   Quantum trajectories for a free electron 
of energy $E=10$ eV. For all the curves, we have chosen 
$x(t=0)= 0$.}
}
\end{figure}


%
%
\begin {equation}
\epsilon^2-{m{\dot{x}}^2 \over 2} \epsilon 
 + {{\hbar}^2\over 8} \left[{3\over 2} 
\left({\ddot{x}\over \dot{x}}\right)^2-{\dot{\ddot{x}} 
\over \dot{x}} 
\right]=0 \; .
\end {equation}
We can check that the general solution of this third order differential 
equation can be written as
\begin {equation}
x(t) = \pm {\hbar \over 2\sqrt{-2m\epsilon} }\ln {\left\vert 
{1 \over a} \tan \left(-{2\epsilon \over \hbar }(t-t_0) \right)
- {b \over a}\right\vert}  \; ,
\end {equation}
where $a$, $b$ and $t_0$ are real integration constants satisfying 
the condition $a \ne 0$. We also observe that this solution can 
also be obtained from (6) after having solved (3).

Relation (17) represents the quantum time equation for a particle 
moving in a constant potential in the classically forbidden region. 
Obviously, there are no nodes and no classical trajectory. The 
velocity is given by
\begin {equation}
\dot{x}(t) = \pm  \sqrt{-{\epsilon\over 2m}} \; 
{1+ \tan^2 [-{2\epsilon (t-t_0)/ \hbar }] \over 
-b +  \tan [-{2\epsilon (t-t_0)/ \hbar }] } \; ,
\end {equation}
or, in term of $x$,
\begin {equation}
\dot{x}(t) =  {1 \over a} \sqrt{-{\epsilon\over 2m}} \; 
\left[ \exp (-2\rho x)+ [a\, \exp (\rho x)+b\, \exp (-\rho x)]^2
     \right] \; .
\end {equation}
where $\rho= \sqrt{-2m \epsilon}/\hbar$. Note that (19) is obtained 
from (17) and (18) by using the plus sign in the RHS of (17). It can 
be straightforwardly obtained from (6) if we take 
$\phi_1=\exp (\rho x)$ and $\phi_2=\exp (-\rho x)$. Of course, 
if we use the minus sign in (17), we can also reproduce the obtained 
result from (6). Relation (18) indicates that if the particle enters 
the classically forbidden region at any time belonging to the interval 
$$
\left \rbrack t_0 -{\pi \hbar \over 2\epsilon }
\left(n-{1 \over 2 }\right),
t_0-{\pi \hbar \over 2\epsilon }
\left(n+{1 \over 2 }\right) \right \rbrack \; ,
$$
its velocity becomes infinite at the time 
$t_0-(2n+1) \pi \hbar /4\epsilon $ 
(we consider a non-relativistic theory). This means that the particle 
takes, at the very most, a time equal to $-\pi \hbar /2\epsilon $ 
before its velocity becomes infinite.

Let us now apply our result for the following rectangular 
potential barrier
\[
V(x) = \left\{ \begin{array}{cc}
               0, & \  \  x<0 \\ [.1in]
               V_0, & \  \  0 \leq x \leq q \\ [.1in]
               0, & \  \ x>q \; .
               \end{array} 
       \right.
\label{eq:ve}
\]
First, we mention that our goal is not to determine the conditions 
for which the particle traverses the barrier. This question 
requires in our point of view further investigations. Our task 
here consists in calculating the time delay in tunneling through 
this barrier and comparing it to earlier results. After we express 
$t$ in terms of $x$ in (17), we easily calculate the time 
necessary for the particle to go from $x=0$ to any point $x$ located 
between 0 and $q$. We get 
\begin {equation}
T(x) \equiv t(x) - t(0) =
     - {\hbar \over 2\epsilon} 
     \left[  \arctan [a \, \exp(2\rho x)+b ]
             - \arctan(a+b) \right]\; .
\end {equation}   
In this relation, $a$ is assumed positive and we have considered 
the positive velocity case. In order to calculate 
the time delay in tunneling through the potential barrier, 
it is sufficient to substitute in (20) $x$ by $q$. For a 
thin barrier $(\rho q \ll 1)$ and a thick one 
$(\rho q \gg 1)$, the above result turns out to be
\begin {equation}
T(q) = {a \over 1 + (a+b)^2}\sqrt{-{2m \over \epsilon}} \; q \; ,
\end {equation}   
and
\begin {equation}
T(q) = -{\hbar \over 2\epsilon} 
      \left[{\pi \over 2} - 
            \arctan (a+b)  \right] \; ,
\end {equation}   
respectively. As for Fletcher's \cite{Fl}
results, in the thin barrier case, the time $T(q)$ is 
proportional to the thickness $q$ and, in the thick barrier one, 
$T(q)$ becomes independent on the thickness. However, 
in contrast with the results of Refs. \cite{Fl, Ha}, ours 
depends only on the difference $|\epsilon| = V_0 - E$ and not 
on $E$ and $V_0$. Furthermore, ours depends also on the 
parameters $a$ and $b$ which themselves depend on the initial 
conditions \cite{BD} and specify the particular microstate 
that we considered, while the ones established in \cite{Fl, Ha} 
are obtained with the use of wave packets.

Now, let us compare our results to those of Floyd \cite{F13}, 
obtained in the context of another formulation of trajectory 
representation. For simplicity, let us choose  
as independent solutions of Schr\"odinger's equation the 
following functions 
\begin {equation}
\phi_1 = \exp(-\rho x)\; , \ \ \ \ \ \  \phi_2 = \exp(\rho x) \; ,
\end {equation}
inside the barrier ($0\leq x \leq q$). Substituting these 
solutions in expression (2) for $S_0$ and using Jacobi's theorem, 
\begin {equation}
t-t_0= {\partial S_0 \over \partial E} \; , 
\end {equation}
as proposed by Floyd \cite{F26}, the expression for $T(x)$ as 
defined in (20) takes the form 
\begin {equation}
T(x)= {2ma \over \hbar \rho}
      {x \; \exp(-2\rho x) \over 1+ [a \; \exp(-2\rho x)+b]^2} \; .
\end {equation}
We indicate that in Floyd's notation, $a$ and $b$ represent 
respectively $b/(ab-c^2/4)^{1/2}$ and $c/2(ab-c^2/4)^{1/2}$. 
First, we remark that for thick barriers $(q \to \infty )$, 
expression (25) leads to $T(q)=0$. This result is different 
from the one obtained by Floyd in \cite{F13} by using another 
couple of solutions of the Schr\"odinger equation. This means 
that the trajectories obtained from Eq. (24) depend on the choice
of mathematical solutions of the Schr\"odinger equation. 
Another problem disclosed by 
relation (25) for thick barriers $(q\to\infty)$ is the fact 
that, after having calculated $dT/dx$, we see that there exists 
a point $x_0$ for which $T(x)$ is a decreasing function for 
$x\geq x_0$. Furthermore, the point $x_0$ is not a turning point 
since $dT/dx$ vanishes at $x_0$ and, then, the velocity is 
infinite. So, we get to the conclusion that the evolution of 
time is reversed and, therefore, the causality of the theory is 
lost. We also indicate that after a tedious calculation, 
even if we use the solutions chosen by Floyd in \cite{F13} to 
calculate $T(x)$, we get to the same conclusion for a set of 
allowed values for $b$. However, in the context of our 
formulation of trajectory representation, there is no interval 
for $x$ for which $T(x)$, as defined in (20), is a decreasing 
function. Furthermore, as we will see in Section 5, 
our equations of motion do not depend on the choice of the 
solutions $\phi_1$ and $\phi_2$.

\vskip0.5\baselineskip
\noindent
{\bf 3.\ \ Linear potential}
\vskip0.5\baselineskip

Let us consider now the linear potential
\begin {equation}
V(x) = gx  \; ,
\end {equation}
where $g$ is a constant which we choose positive. First, remark  
that the Schr\"o\-dinger equation can be written in the form 
of Airy equation
\begin {equation}
{d^2\phi \over dy^2 }-y\phi (y) = 0 \; ,
\end {equation}
where
\begin {equation}
y = \left({ 2m \over \hbar^2 g^2}\right)^{1/3}(gx - E) \; .
\end {equation}
The series method allow us to get for Eq. (27) two real 
independent solutions which can be related to Airy 
functions $Ai$ and $Bi$ as
\begin {equation}
\phi_1(y)=Ai(y)+{1\over \sqrt{3}}Bi(y)\; ,
\end {equation}
\begin {equation}
\phi_2(y) = \sqrt{3} Ai(y) - Bi(y)\; .
\end {equation}
%
%
\begin{figure}
\def\put(#1,#2)#3{\leavevmode\rlap{\hskip#1\unitlength\raise#2\unitlength\hbox{#3}}}
\centerline{
\vbox{\hsize=12.5cm
\setlength{\unitlength}{1truecm}
\put(0,0){\epsfxsize=12cm \epsfbox{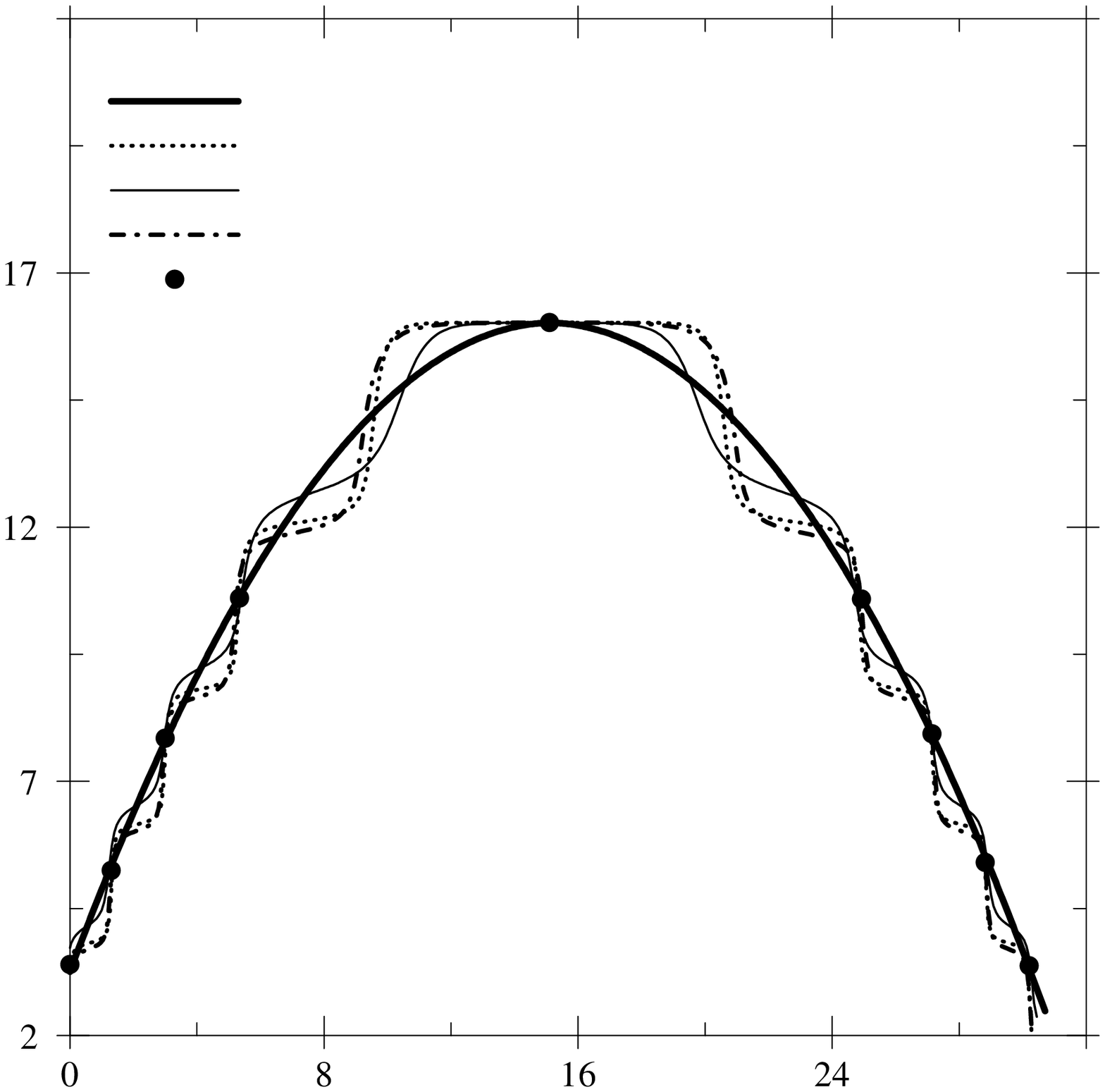}}%
\put(10.2,0.8) {$t$ ($\times 10^{-16}$ s)}%
\put(0.5,11.7) {$x$ ($\times 10^{-10}$ m)}%
\put(3.15,10.52) {\footnotesize  Classical trajectory }%
\put(3.01,10.12) { \footnotesize $ a=10$, $b=\sqrt{3}$ }%
\put(3.01,9.66) { \footnotesize $ a=7$, $b= -1$  }%
\put(3.01,9.20) { \footnotesize $ a=5$, $b=\sqrt{3}$ }%
\put(3.01,8.71) { \footnotesize Nodes}%
}}
\centerline {
\vbox 
{ \hsize=12cm\noindent 
\small Fig. 2: Quantum trajectories for an electron 
of energy $E=10$ eV moving in a linear potential $V(x)=gx$ 
($g=10^{-9}$ kg m s$^{-2}$) in the classically allowed region. 
For all the curves, we have chosen 
$x(t=0)=3.25405 \times 10^{-10}$ m. 
The maximum of the curves is located at 
 $(t=14.47545 \times 10^{-16}$ s, $x=16.02189 \times 10^{-10}$ m).  }
}
\end{figure}
%
%
%
The equation of motion is obtained by substituting 
(29) and (30) in (6) 
\begin {eqnarray}
{dx \over dt} = \pm {2(E-gx) \over (2mg\hbar)^{1/3}aW }
\left[ 
\left(a^2+3b^2+2\sqrt{3}ab+3 \right)Ai^2(y)
\right. \hskip25mm&& \nonumber\\
\left. + 2\left({a^2 \over \sqrt{3}}-\sqrt{3}b^2 -
\sqrt{3} \right)Ai(y)Bi(y) + 
\left({a^2 \over 3}+b^2 - {2ab\over\sqrt{3}} +1 \right)Bi^2(y)
\right] \; ,
\end {eqnarray}
where 
$$ 
W={d \phi_1 \over dy }\phi_2 -
{d \phi_2 \over dy }\phi_1
=2\left[{ dBi \ \over dy } Ai -
{dAi \over dy } Bi \right] = {2 \over \pi} \; 
$$
is the Wronskian of $\phi_1$ and $\phi_2$. Eq. (31) is 
valid both in the classically allowed case and the forbidden one. 
It is a first order differential equation in which we see the 
presence of three integration constants $E$, $a$ and $b$. Since 
it does not have an exact solution, we have appealed to numerical 
methods. In Fig. 2, we have plotted from (31) in $(t,x)$ plane 
some trajectories corresponding to different values of $a$ and 
$b$ in the classically allowed case $(y\leq 0)$. The considered 
system is an electron of energy $E=10$ eV and we have chosen 
$g=10^{-9}$ kg m s$^{-2}$. From the classical analogue of 
Eq. (31), given by 
\begin {equation}
{dx \over dt} = \pm \sqrt{{2 \over m}(E-gx)} \; ,
\end {equation}
%
%
%
%
\begin{figure}
\def\put(#1,#2)#3{\leavevmode\rlap{\hskip#1\unitlength\raise#2\unitlength\hbox{#3}}}
\centerline{
\vbox{\hsize=10.5cm
\setlength{\unitlength}{1truecm}
\put(0,0){\epsfxsize=10cm \epsfbox{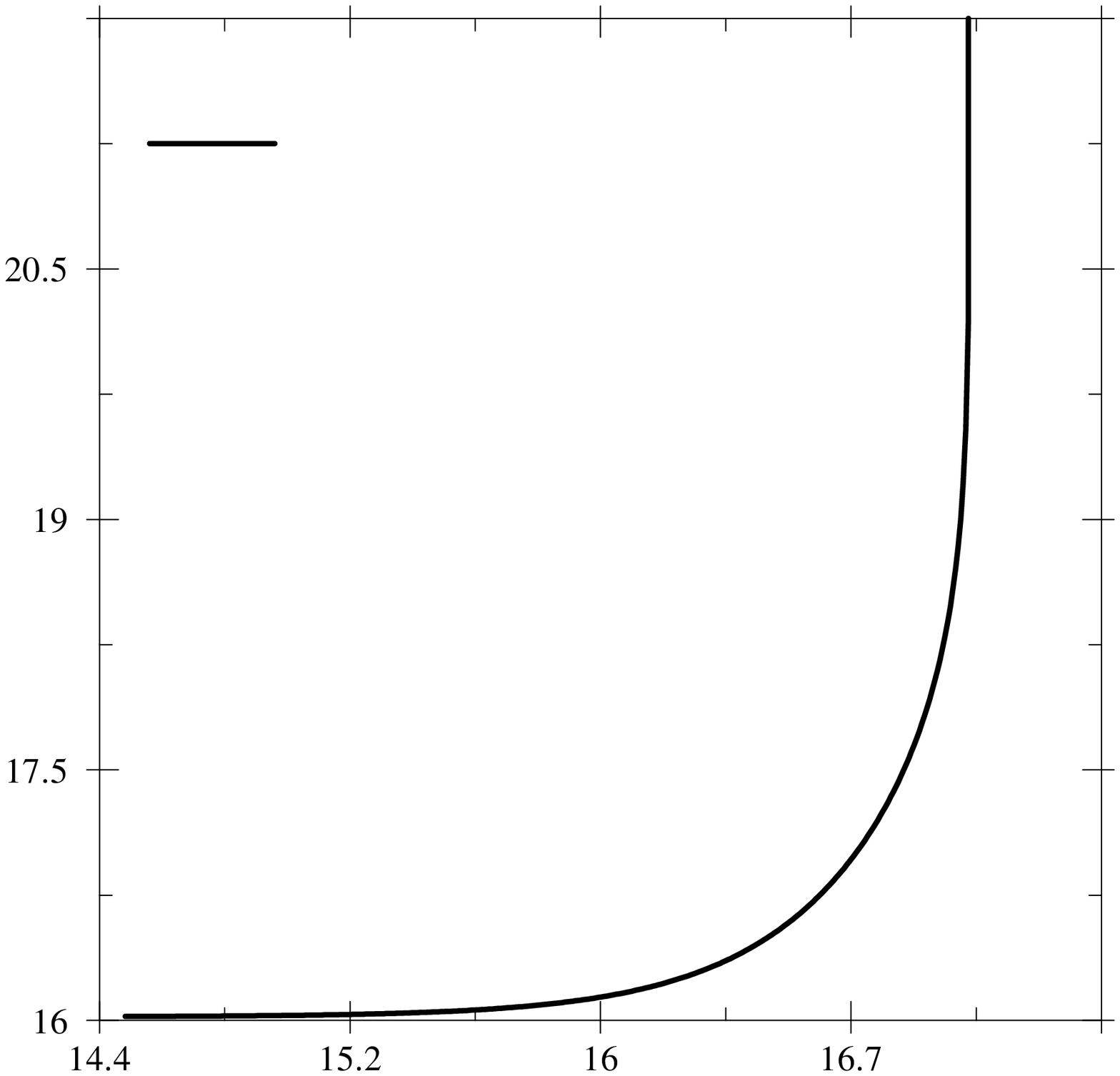}}%
\put(8.5,0.6) {$t$ ($\times 10^{-16}$ s)}%
\put(0.2,9.75) {$x$ ($\times 10^{-10}$ m)}%
\put(2.6,8.4) { \footnotesize $ a=10 ,\  b=1/\sqrt{3}$}%
}}
\centerline {
\vbox 
{ \hsize=12cm\noindent 
\small Fig. 3: Quantum trajectory for an electron of 
energy $E=10$ eV moving in a linear potential $V(x)=gx$ 
($g=10^{-9}$ kg m s$^{-2}$) in the classically forbidden region. 
We have chosen $a=10$, $b=1/\sqrt{3}$ and $x(t=14.47546 
\times 10^{-16}$ s) $=16.02190 \times 10^{-10}$ m.   }
}
\end{figure}
%
we have plotted in the same figure the classical trajectory.
As in the constant potential case, the quantum trajectories 
oscillate about the classical one. We observe that all 
trajectories, even the classical one, pass through some 
points constituting nodes. In particular, for all possible 
trajectories, the velocity has a vanishing value at $y=0$. 
As explained in Section 1, we indicate that for 
the trajectories plotted in Fig. 2, we have used the plus 
sign in the RHS of (31) in the domain where $\dot{x}>0$ and 
the minus sign in the domain where $\dot{x}<0$. This is also  
the case in (32) for the classical trajectory.

In contrast with the constant potential case, the distance 
between two adjacent nodes is not constant. We remark that the 
two intervals starting from the node where the velocity 
vanishes (at $y=0$) are the most long ones. The length of the 
intervals decreases gradually as the velocity increases along 
the trajectories.  We will explain this observation in Section  
5 and show that this length is proportional to $\hbar$. This  
means that in the classical limit $\hbar \to 0$, the 
adjacent nodes become infinitely close, and as we will see in 
Section 5, the quantum trajectories tend to be identical to 
the classical one.

We remark also that, in contrast with the constant potential 
case, there are no particular values for $a$ and $b$ for  
which the quantum trajectories reduce to the classical one. 
In fact, the RHS of (31) can be developed as a 
power series in $y$ while the RHS 
of (32) is proportional to $\sqrt{-y}$. However, it 
is peculiar to observe that for the particular values 
$a=2$ and $b=-1/\sqrt{3}$, the quantum trajectory for 
$y<0$ is quasi identical to the classical 
one. This result is in agreement with the fact that 
$Ai^2(y)+Bi^2(y)$ acts like $1/\sqrt{-y}$.

Now, let us consider the classically forbidden region $(y>0)$. 
As in the constant potential case, our investigations do not 
concern the conditions for which the particle enters  
this region. We suppose only that the particle is present 
in this region and we determine its trajectory from the 
equation of motion (31) by appealing to numerical methods.
As an example, we have considered in Fig. 3 an electron with 
energy $E=10$ eV for the particular values $a= 10$ and 
$b=1/\sqrt{3}$. We see that as soon as the particle enters 
this region, its velocity increases quickly. We have 
checked that there are no nodes.

\vskip0.5\baselineskip
\noindent
{\bf 4.\ \ Harmonic oscillator}
\vskip0.5\baselineskip

Without appealing to the usual axiomatic interpretation of 
the wave function, Faraggi and Matone showed \cite{FM3,FM4} 
that energy quantization is a consequence of the equivalence 
postulate \cite{FM1,FM2,FM3}. The case of the harmonic 
oscillator is particularly studied in Ref. \cite{FM3}. 
In one dimension, the potential is given by 
\begin {equation}
V(x)={1 \over 2}m\omega^2x^2 \; .
\end {equation}
Let us begin by the fundamental state for which the physical 
wave function, up to a constant factor, is given by
\begin {equation}
\phi_2(x)=\exp(-\alpha x^2) \; ,
\end {equation}
where $\alpha = m \omega /2 \hbar$. The relationship between the 
corresponding energy and the frequency is $E_0 = \hbar \omega /2$. 
A second independent solution of the Schr\"odinger equation 
can be obtained by using the fact that the Wronskian is constant
\begin {equation}
\phi_1(x)=\exp(-\alpha x^2) 
\int_{x_0}^{x} \exp(2 \alpha q^2) \; dq \; .
\end {equation}
Here, we have chosen the Wronskian $W(\phi_1,\phi_2) = 
\phi_2 \; d \phi_1/dx - \phi_1 \; d \phi_2 /dx = 1$. 
Note that the lower boundary $x_0$ of the 
integral in (35) can be arbitrary chosen. Thus, in what 
follows, we set $x_0=0$ and, then, $\phi_1(x)$ represents 
the Dawsons integral. Substituting (34) and (35) in (6), 
the equation of motion takes the form
%
\begin {equation}
{dx \over dt}= \pm {2E_0 \over \hbar a}(1-2\alpha x^2) 
\exp (-2\alpha x^2) 
\left[ 
       1 + \left( a \int_{0}^{x} \exp(2 \alpha q^2) \; dq +b\right)^2
       \right] \; .
\end {equation}

Again, there is no exact solution for $x$. Numerical methods 
allow us to plot some trajectories corresponding to different 
values of $a$ and $b$. In Fig. 4, we have considered in the 
classically allowed region $(|x| \leq x_{M_0})$ the motion of an 
electron of energy $E_0=10$ eV over one period. Here, $x_{M_0}$ 
represents the corresponding classical amplitude
\begin {equation}
x_{M_0}=\sqrt{{2E_0\over m\omega^2}}={\hbar \over \sqrt{2mE_0}} \; .
\end {equation}   
%
%

%
\begin{figure}
\def\put(#1,#2)#3{\leavevmode\rlap{\hskip#1\unitlength\raise#2\unitlength\hbox{#3}}}
\centerline{
\vbox{\hsize=10.5cm
\setlength{\unitlength}{1truecm}
\put(0,0){\epsfxsize=10.cm \epsfbox{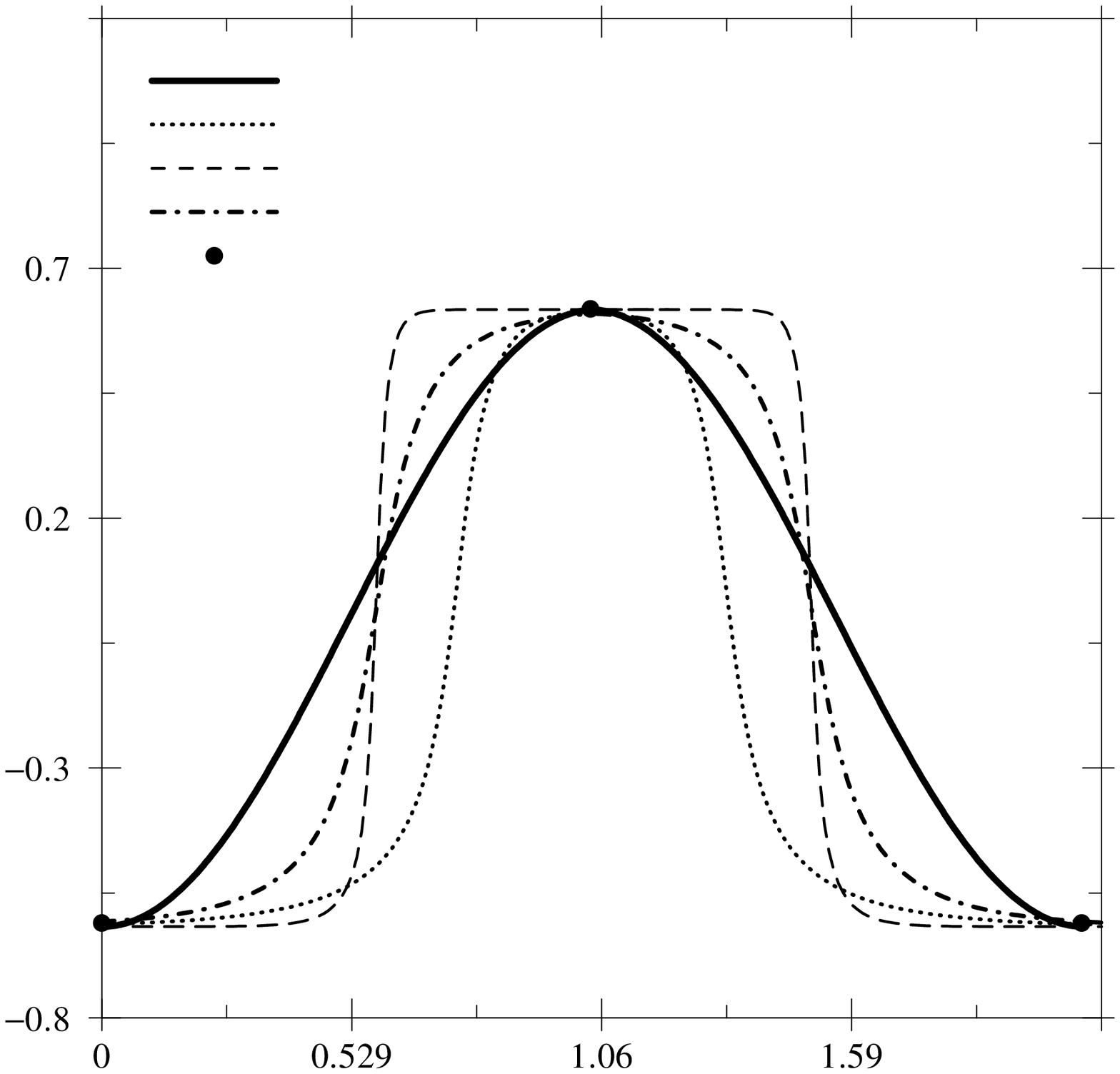}}%
\put(8.5,0.6) {$t$ ($\times 10^{-16}$ s)}%
\put(0.2,9.75) {$x$ ($\times 10^{-10}$ m)}%
\put(2.77,8.95) {\footnotesize  Classical trajectory }%
\put(2.62,8.60) { \footnotesize $a=8 \times 10^{9}$, $b=1$ }%
\put(2.62,8.25) { \footnotesize  $a=6 \times 10^{10}$, $b=2$ }%
\put(2.62,7.85) { \footnotesize $a=9 \times 10^{9}$, $b=0.2$ }%
\put(2.62,7.40) { \footnotesize Nodes}%
}}
\centerline {
\vbox 
{ \hsize=12cm\noindent 
\small Fig. 4: Quantum trajectories for the fundamental state 
of energy $E_0=10$ eV of a harmonic oscillator in the classically 
allowed region. For all the curves, we have chosen 
$x(t=0)=-x_{M_0}= -0.61725 \times 10^{-10}$ m. The first 
maximum of the curves is located at 
$(t=1.04200 \times 10^{-16}$ s, $x=x_{M_0}=0.61725 \times 10^{-10}$ m).   }
}
\end{figure}
%
%

We observe the presence of nodes in the $(t,x)$ plane at the points 
$x=-x_{M_0}$ and $x=x_{M_0}$ corresponding to the vanishing values 
of the velocity. We notice that, even if we impose a node by 
choosing for trajectories the same initial condition 
$x(t=0)=x_0$ at any point inside the interval 
$\rbrack -x_{M_0},x_{M_0} \lbrack$, all the following nodes in the 
$(t,x)$ plane will be at the points $x=\pm x_{M_0}$ where the 
velocity vanishes. On the other hand, we indicate that at the 
half-periods where the velocity is positive (negative), we have 
used the plus (minus) sign in the RHS of (36). In the classical 
limit $\hbar \to 0$, the oscillator becomes a point at rest 
because the classical amplitude vanishes.

The classical analogue of Eq. (36) is 
\begin {equation}
{dx \over dt}= \pm \sqrt{{2 \over m}(E_0-{1 \over 2}m\omega^2x^2}) \; .
\end {equation}
As in the linear potential case, there are no particular values 
for $a$ and $b$ for which the quantum equation (36) reduces to 
the classical equation (38). However, it is peculiar to observe 
that for $a=10^{10}$ and $b=0$, the quantum trajectory 
plotted from (36) resembles the classical one.


\begin{figure}
\def\put(#1,#2)#3{\leavevmode\rlap{\hskip#1\unitlength\raise#2\unitlength\hbox{#3}}}
\centerline{
\vbox{\hsize=10.5cm
\setlength{\unitlength}{1truecm}
\put(0,0){\epsfxsize=10cm \epsfbox{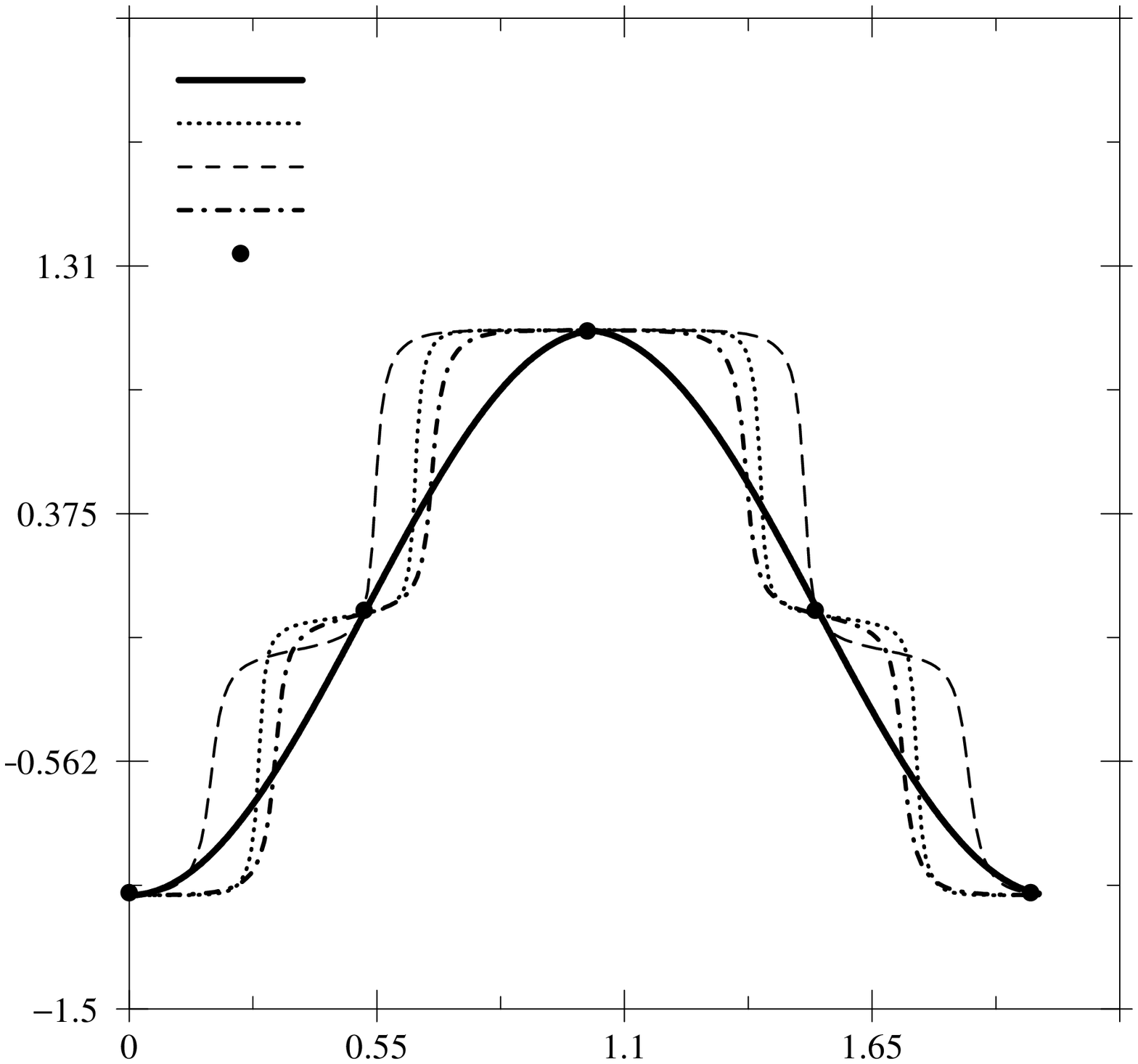}}%
\put(8.5,0.6) {$t$ ($\times 10^{-16}$ s)}%
\put(0.2,9.75) {$x$ ($\times 10^{-10}$ m)}%
\put(2.75,8.94) {\footnotesize Classical trajectory }%
\put(2.65,8.57) { \footnotesize $ a=4 \times 10^{8} ,\ b=0.5$}%
\put(2.65,8.17) { \footnotesize $ a=5 \times 10^{9} ,\ b=0.4$}%
\put(2.65,7.8) { \footnotesize $ a=5 \times 10^{8}  ,\  b=0$}%
\put(2.65,7.4) { \footnotesize Nodes}%
}}
\centerline {
\vbox 
{ \hsize=12cm\noindent 
\small Fig. 5: Quantum trajectories for the first excited  
state of energy $E_1=30$ eV of a harmonic oscillator  in the 
classically allowed region. For all the curves, we have chosen 
$x(t=0)=-x_{M1}=-1.06911 \times 10^{-10}$ m. 
The first maximum of the curves is located at
$(t=1.04200 \times 10^{-16}$ s, $x=x_{M1}=1.06911 \times 10^{-10}$ m).
}}
\end{figure}


Now, consider the first excited state. The physical solution of 
Schr\"odinger's equation is
\begin {equation}
\phi_2(x)=x\exp(-\alpha x^2) \; .
\end {equation}
The relationship between the corresponding energy and the 
frequency is $E_1 = 3 \hbar \omega /2$. It follows that the 
amplitude of the corresponding classical oscillator is 
$x_{M1} = 3 \hbar / \sqrt{2mE_1}$. Its ratio with the 
corresponding amplitude of the fundamental state is 
$\sqrt{3}$. A second independent solution  
is obtained by using the fact that the Wronskian is constant
\begin {equation}
\phi_1(x)=x\exp(-\alpha x^2) 
\int_{0}^{x} {\exp(2 \alpha q^2) \over q^2}  dq \; .
\end {equation}
Here, we have chosen the Wronskian $W(\phi_1,\phi_2) = 1$. 
As in the fundamental state case, we substitute (39) and (40) 
in (6) to obtain the quantum equation of motion from which 
we plot some trajectories (Fig. 5) for different 
values of $a$ and $b$. The value of the energy, $E_1=30$ eV, 
that we take is three times that of the fundamental state. 
We remark that we have an additional 
node for every half-period of the oscillator motion compared 
to the fundamental state case. As we will explain in 
the next section, this additional node is a consequence of the 
zero of the function $\phi_2(x)$ given by (39). 


\begin{figure}
\def\put(#1,#2)#3{\leavevmode\rlap{\hskip#1\unitlength\raise#2\unitlength\hbox{#3}}}
\centerline{
\vbox{\hsize=10.5cm
\setlength{\unitlength}{1truecm}
\put(0,0){\epsfxsize=10cm \epsfbox{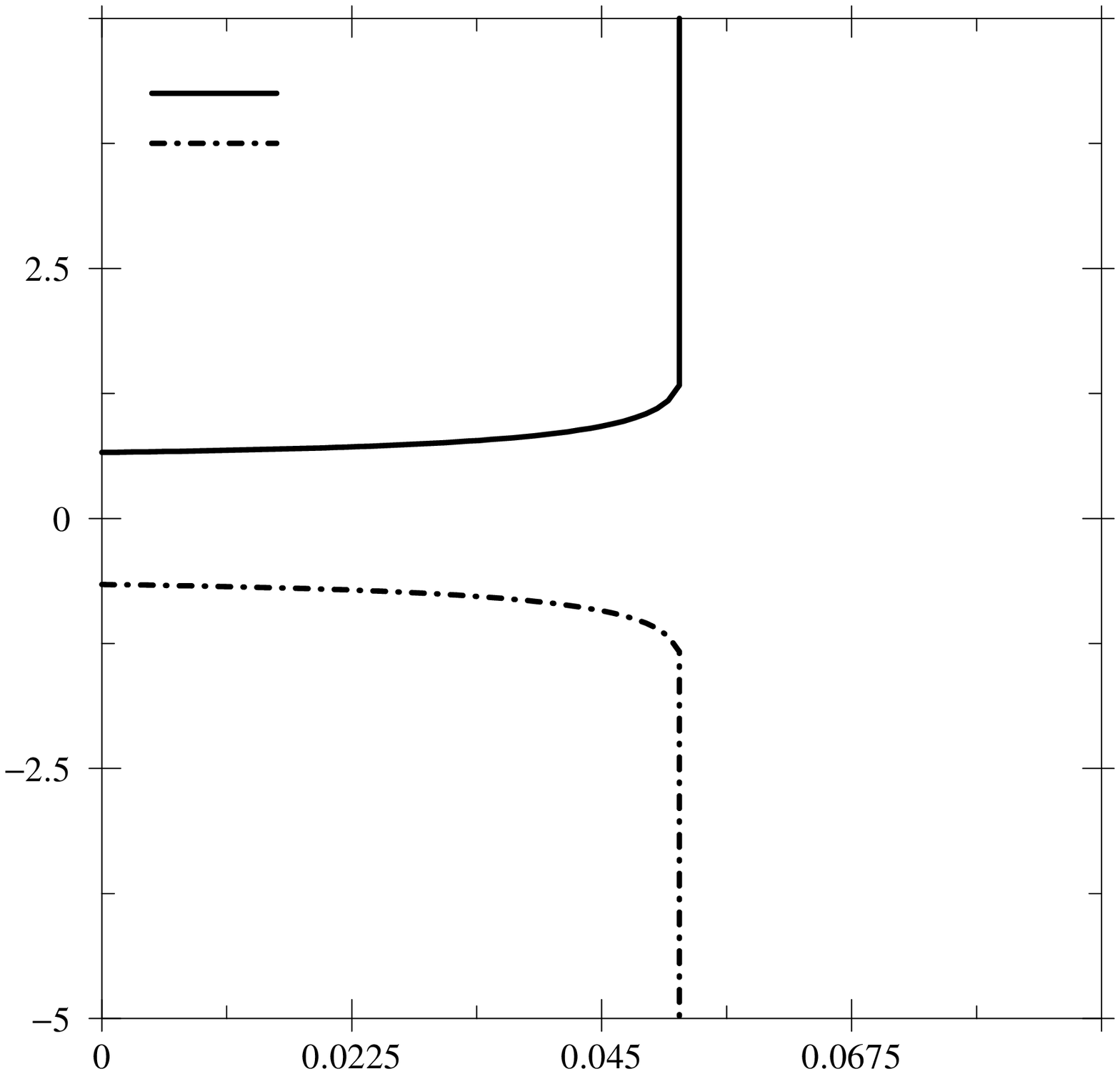}}%
\put(8.5,0.6) {$t$ ($\times 10^{-16}$ s)}%
\put(0.2,9.75) {$x$ ($\times 10^{-10}$ m)}%
\put(2.8,8.82) {\footnotesize $ a=8 \times 10^{10} ,\ b=1 $}%
\put(2.7,8.4) { \footnotesize $ a=8 \times 10^{10} ,\ b=-1 $}%
}}
\centerline {
\vbox 
{ \hsize=12cm\noindent 
\small Fig. 6: Quantum trajectories for the fundamental
state of energy $E_0=10$ eV of a harmonic oscillator in 
the classically forbidden region. For the curve plotted in 
the domain where  $x>x_{M_0}$, we have chosen $x(t=0)=0.61726 
\times 10^{-10} $ m and for the one plotted in the domain 
where $x<-x_{M_0}$, $x(t=0)=-0.61726  \times 10^{-10}$ m. 
 }
}
\end{figure}
%

Concerning the classical forbidden case, both for the fundamental 
and the first excited states, we remark that as soon as the 
particle enters this region, the velocity increases quickly. 
The nodes do not appear. In Fig. 6, we plotted $x(t)$ 
for the fundamental state with $a=8 \times 10^{10}$ and $b=1$ 
in the case where $x>x_{M_0}$ and with $a=8 \times 10^{10}$ 
and $b=-1$ in the case where $x<-x_{M_0}$. 


\vskip0.5\baselineskip
\noindent
{\bf 5.\ \ General potential and de Broglie's wavelength}
\vskip0.5\baselineskip

Concerning the classically forbidden region, we remark that for all 
the potentials considered here, the velocity increases quickly.
We think that this is the case for any another potential. 
This rapid divergence seems to be in agreement with the predictions 
of Copenhagen School. In fact, in a natural way, we can assume 
the existence of a link between the time the particle stays in an 
interval and the probability of finding this particle in it. The 
rapid divergence may then be explained by the fact that the 
probability density decreases rapidly in the classically forbidden 
regions. As an example, in the harmonic oscillator case, the 
probability density decreases as $\exp(-2 \alpha x^2)$.
 
Let us now consider the classically allowed region. 
The general idea which emerges from the 
previous sections is that, to each classical trajectory, 
we can associate a family of quantum trajectories which can be 
specified by the different values of the non-classical integration 
constants $a$ and $b$. These quantum trajectories oscillate about 
their corresponding classical one which contains some points 
called nodes through which pass all the trajectories of the family. 
Since the nodes are obtained in the $(t,x)$ plane, the time 
the particle takes to go from one node to another is the same 
for all possible trajectories, even for the classical one. 

In the constant potential case, the existence of these nodes is 
shown with an analytical method. We have seen that they are 
strongly linked to the zeros of the function appearing 
in  the denominator of the expression of the reduced 
action $S_0$. We can also check graphically that the obtained 
nodes in the linear potential and the harmonic oscillator 
cases correspond to turning points or to zeros of the 
Schr\"odinger solution used in the denominator appearing in the 
expression of $S_0$. This strongly suggests that for any potential, 
we will obtain nodes in these particular points. Furthermore,
from Eq.(6), we see that for any potential, the velocity does 
not depend on the values of $b$ at the zeros of $\phi_2$.

On the other hand, in the constant potential case, we showed 
that the distance on the $x$ axis between two adjacent nodes is 
a constant given by expression (13).  This distance  is related to 
de Broglie's wavelength 
\begin {equation}
\lambda = {h \over p}
\end {equation}
by
\begin {equation}
\Delta x_n = {\lambda \over 2} \; .
\end {equation}
In (41), $p$ is the classical momentum
\begin {equation}
p = mv\; .
\end {equation}
Note that $v$ can be considered as the classical velocity or as the 
mean velocity of any quantum trajectory between the two nodes. In 
fact, by using (12) and (13), we have  
\begin {equation}
v = {\Delta x_n \over \Delta t_n } = \sqrt{{2 \epsilon \over m}}\; .
\end {equation}
It is important to observe that $p$ also represents the average 
of the quantum conjugate momentum along one interval separating 
two nodes. In fact, taking into account (10), (11) and (13), and 
using (2) to determine $S_0$ in the case where 
$V(x) = V_0$, we can deduce that
\begin {equation}
\left<{\partial S_0 \over \partial x }\right> \equiv 
{1 \over \Delta x_n} \int_{x(t_n)}^{x(t_{n+1})} 
{\partial S_0 \over \partial x } \; dx 
= {S_0(x(t_{n+1}))-S_0(x(t_n)) \over \Delta x_n} =
\sqrt{2m \epsilon} \; ,
\end {equation}
which is equal to $p$ with the use of (43) and (44). This 
result suggests strongly and in a natural way, that for any 
potential we define a new wavelength associated to any interval 
between two adjacent nodes as in (41) except that $p$ must be 
substituted by 
\begin {equation}
p = \left<{\partial S_0 \over \partial x }\right> \; . 
\end {equation}
Therefore, by using expression (2) for $S_0$ to average 
$\partial S_0 / \partial x$ between two adjacent zeros of 
$\phi_2$, we obtain for any potential
\begin {equation}
p = {\pi \hbar \over \Delta x} \; ,
\end {equation}
$\Delta x$ being the length between the two zeros. 
It also represents the length between the two corresponding 
nodes. Substituting (47) in (41), we obtain 
\begin {equation}
\Delta x = {\lambda \over 2} \; ,
\end {equation}
as it is for the constant potential case, Eq. (42). This relation 
gives the link between the length separating adjacent nodes 
and the new wavelength as defined by (41) and (46). We stress 
that we do not associate any wave to our particle motion 
but we just keep the terminology introduced by de Broglie. 

Taking into account (41) and (48), the previous conclusion implies 
that the distance between adjacent nodes is also proportional 
to $\hbar$, as it is in the constant potential case. We deduce 
therefore that for any potential in the classical limit 
$\hbar \to 0$, the adjacent nodes become infinitely close.
As in the constant potential case, this finding implies that the 
quantum trajectories tend to be identical to their corresponding 
classical one. In fact, since

- the particular expression for $x(t)$ is not used in our 
reasoning for the constant potential in Section 2,

- Eq. (6) indicates that the function $x(t)$ is monotonous 
 between two adjacent nodes for any potential, 

- in the classical limit $(\hbar \to 0)$, the classical trajectory 
between two adjacent nodes can be assimilated to an infinitesimal 
straight segment,

\noindent
our reasoning in Section 2 can be easily generalized 
for any potential. Now, we can assert that, for any 
potential, the classical limit $(\hbar \to 0)$ of 
any quantum trajectory is the classical trajectory. 
This conclusion is compatible with the fact that 
the quantum equations of motion, Eq. (5), the FIQNL 
(Eq. (7)) and  even the QSHJE (Eq. (1)), become all identical to 
their corresponding classical equations in the limit $\hbar \to 0$. 
It will not be logical if the quantum time equations do not have as 
a limit the classical equations when $\hbar \to 0$, while the 
quantum equations of motion have as a limit the classical ones. 

An important quantity to determine is the Ermakov invariant 
\cite{E20,L18}. In the context of Schr\"odinger's equation, 
this invariant has been first introduced by Floyd \cite{F214} 
and later written by Faraggi-Matone \cite{FM3} as 
\begin {equation}
I = {1\over \sqrt{2m}}
    \left[{\partial S_0 \over \partial x}\psi_E^2 
     +\hbar^2 \left[
     {1 \over 2} \left({\partial S_0 \over \partial x}\right)^{-3/2}
     {\partial^2 S_0 \over \partial x^2}\psi_E 
     + \left({\partial S_0 \over \partial x}\right)^{-1/2}
     {\partial \psi_E \over \partial x}
     \right]^2 
     \right] \; , 
\end {equation}
where $\psi_E$ is the physical solution of Schr\"odinger's 
equation. Of course $\psi_E$ can be written as
\begin {equation}
\psi_E = \alpha \phi_1 + \beta \phi_2 \; , 
\end {equation}
where $\alpha$ and $\beta$ are complex constants. With the 
use of (4) and (50), we can show that (49) leads to
\begin {equation}
I =  {\hbar W \over a \sqrt{2m}}\left[\alpha^2 + 
     (a\beta -b\alpha)^2\right]  \; . 
\end {equation}
It is clear that $I$ is an invariant.

Another important question which we must investigate concerns 
the link between the nodes and the zeros of the 
function $\phi_2$: do quantum trajectories depend on the 
choice of $\phi_2$? In other words, we are afraid that the 
mathematical choices may affect our physics results.

In order to answer this crucial question, let us consider a new set 
of real solutions of Schr\"odinger's equation, Eq. (3),
\begin {equation}
\theta_1 = \mu \phi_1 + \nu \phi_2 \; , 
\end {equation}
\begin {equation}
\theta_2 = \alpha \phi_1 + \beta \phi_2 \; .
\end {equation}
We suppose that the real parameters $(\mu, \nu, \alpha, \beta)$ 
satisfy the condition $\mu \beta - \nu \alpha \not= 0$ in such a 
way as to guarantee the fact that $\theta_1$ and $\theta_2$ must be 
independent. Let us look for the existence of a couple of 
parameters $(\tilde{a},\tilde{b})$ with which  the reduced action 
takes the form 
\begin {equation}
S_0=\hbar \ \arctan {\left [ \tilde{a}  {\theta_1 
\over \theta_2 } + \tilde{b} \right ]} +\hbar \tilde{l} \; ,
\end {equation}
as in (2), and from which  we deduce the same equation of motion, 
Eq. (6). For this purpose, let us apply the fundamental relation (5) 
in which we substitute $S_0$ by expression (54). Taking into 
account relations (52) and (53), we obtain
\begin {eqnarray}
{dx \over dt}= \pm { 2[E-V(x)] \over \hbar W}
\left[ {\mu^2 \tilde{a}^2 +2\mu \alpha \tilde{a}\tilde{b} +
\alpha^2 (1+\tilde{b}^2) \over (\mu \beta - \nu \alpha)
\tilde{a} }\phi_1^2 \right. \hskip30mm&& \nonumber\\
+2 {\mu \nu \tilde{a}^2 + (\mu \beta+\nu \alpha ) 
\tilde{a}\tilde{b} +\alpha \beta (1+\tilde{b}^2)
\over (\mu \beta - \nu \alpha)\tilde{a} } \phi_1\phi_2 
\hskip20mm&& \nonumber\\
+ \left. {\nu^2 \tilde{a}^2 + 2 \beta \nu \tilde{a} \tilde{b} +
\beta^2(1+\tilde{b}^2) \over (\mu \beta - \nu \alpha)
\tilde{a} } \phi_2^2 \right] \; ,
\end {eqnarray}
where we have used the fact that the Wronskian $\tilde{W}$ 
of $(\theta_1, \theta_2)$ is related to the one of 
$(\phi_1,\phi_2)$, $W$, by 
$\tilde{W} = (\mu \beta - \nu \alpha) W$. 
Equation of motion (55) is identical to (6) if and only if 
\begin {equation}
a = {\mu^2 \tilde{a}^2 +2\mu \alpha \tilde{a}\tilde{b} +
\alpha^2 (1+\tilde{b}^2) \over (\mu \beta - \nu \alpha)
\tilde{a} } \; ,
\end {equation}
\begin {equation}
b = {\mu \nu \tilde{a}^2 + (\mu \beta+\nu \alpha ) 
\tilde{a}\tilde{b} +\alpha \beta (1+\tilde{b}^2)
\over (\mu \beta - \nu \alpha)\tilde{a} } \; ,
\end {equation}
\begin {equation}
{1+b^2 \over a } = {\nu^2 \tilde{a}^2 + 2 \beta \nu \tilde{a} \tilde{b} +
\beta^2(1+\tilde{b}^2) \over (\mu \beta - \nu \alpha)
\tilde{a} }  \; .
\end {equation}
The parameters $\tilde{a}$ and  $\tilde{b}$ can be determined 
from (56) and (57). On the other hand, if we substitute expressions 
(56) and (57) for $a$ and $b$ in (58), we find that (58) 
represents an identity. This means that (58) is compatible with 
(56) and (57). Therefore, for any couple $(\theta_1, \theta_2)$ 
defined by $(\mu, \nu, \alpha, \beta)$, it is always possible 
to get parameters $(\tilde{a}, \tilde{b})$ with which we reproduce 
the same quantum motion as the one given by (6) which  we 
deduce from the reduced action (2). In conclusion, the 
mathematical choices of $(\phi_1, \phi_2)$  do not affect the 
physics results.

 
\vskip\baselineskip
\noindent
{\bf REFERENCES}

\end {document}